\documentclass[12pt]{article}
\usepackage[utf8]{inputenc}
\usepackage{amsmath,amssymb,amsfonts,mathtools,amsthm}
\usepackage{bm,bbm}
\usepackage{booktabs}
\usepackage{color}
\usepackage[dvips,letterpaper]{geometry}
\usepackage{epsfig}
\usepackage{fullpage}
\usepackage{graphicx,psfrag,epsf}
\usepackage{indentfirst}
\usepackage{lineno}
\usepackage{natbib}
\usepackage{setspace}
\usepackage{subfigure}
\usepackage{times}
\usepackage{url}
\usepackage{verbatim}
\usepackage{multirow}
\usepackage[table,xcdraw]{xcolor}
\usepackage{listings}
\usepackage{tikz}
\usetikzlibrary{matrix,shapes,arrows,positioning}
\usepackage{epstopdf} 
\usepackage{caption}                
\usepackage{float} 
\floatstyle{plaintop} 
\restylefloat{table} 
\usepackage[colorlinks, linkcolor=black, citecolor= blue]{hyperref} 
\usepackage{algorithm,algcompatible,algpseudocode}
\usepackage{setspace}

\def \build#1#2#3{\mathrel{\mathop{#1}\limits^{#2}_{#3}}}

\title{\textbf{\LARGE Parametric quantile autoregressive conditional duration models with application to intraday value-at-risk}}

\author{\normalsize
\bf 
Helton Saulo$^{1,2}$, Suvra Pal$^{2}$, Rubens Souza$^{1}$, Roberto Vila$^{1}$, Alan Dasilva$^{3}$, \\
 {\small $^{1}$Department of Statistics, Universidade de Bras\'{i}lia, Bras\'{i}lia, Brazil}\\[-0.15cm]
 {\small $^{2}$Department of Mathematics, University of Texas at Arlington, Arlington, TX, USA}\\[-0.05cm]
 {\small $^{3}$Institute of Mathematics and Statistics, Universidade de S\~{a}o Paulo, S\~{a}o Paulo, Brazil}\\[-0.05cm]
}

\date{} 

\begin{document}
\maketitle
\begin{center}
\begin{minipage}{0.9\linewidth}
\begin{spacing}{1}
\paragraph{Abstract:}
	
The modeling of high-frequency data that qualify financial asset transactions has been an area of relevant interest among statisticians and econometricians -- above all, the analysis of time series of financial durations. Autoregressive conditional duration (ACD) models have been the main tool for modeling financial transaction data, where duration is usually defined as the time interval between two successive events. These models are usually specified in terms of a time-varying mean (or median) conditional duration. In this paper, a new extension of ACD models is proposed which is built on the basis of log-symmetric distributions reparametrized by their quantile. The proposed quantile log-symmetric conditional duration autoregressive model allows us to model different percentiles instead of the traditionally used conditional mean (or median) duration. We carry out an in-depth study of theoretical properties and practical issues, such as parameter estimation using maximum likelihood method and diagnostic analysis based on residuals.  A detailed Monte Carlo simulation study is also carried out to evaluate the performance of the proposed models and estimation method in retrieving the true parameter values as well as to evaluate a form of residuals. Finally, the proposed class of models is applied to a price duration data set and then used to derive a semi-parametric intraday value-at-risk (IVaR) model.

\vspace{0.5cm}

\end{spacing}
\end{minipage}
\end{center}

\onehalfspacing

\section{Introduction} \label{sec:01}

The increasing availability of large volume of data, in different formats and at high frequency, combined with technological advances in tools for storing and processing such data have created a new problem -- the so-called Big Data problem. Records of all financial transactions -- trades, quotes, price changes -- are available on many stock exchanges. An inherent characteristic of such data is that they are irregularly spaced in time. Furthermore, such data contains useful information about market activities and has a number of unique characteristics, such as: (C1) intraday seasonality, wherein activity is higher at the beginning and at the close than during the middle of the trading day; (C2) positive skewness of data distribution; (C3) a hazard rate with an inverse bathtub shape (unimodal); and (C4) a large number of observations (in some cases the transformation into durations largely reduces the number of observations); see \cite{pk:08}, \cite{bv:19}, \cite{engle:00}, \cite{lslm:14}, \cite{b:10} and \cite{sjla:19,sbv:23}.


There are several approaches to handling this data resource. The primary class of models for these analyses has been the autoregressive conditional duration (ACD) model, introduced by \cite{er:98}. These models are commonly used to capture the structure of clusters and can be seen as a counterpart to the autoregressive models with heteroscedasticity (ARCH), proposed by \cite{engle:82}, and its extension, the generalized ARCH models (GARCH), proposed by \cite{blv:86}. Since then, there have been many generalizations to the original ACD model; see, \cite{l:99}, \cite{gm:00}, \cite{bg:00}, \cite{fg:06},
\cite{c:07}, \cite{dz:06}, \cite{zrt:01}, \cite{pk:08}, \cite{bv:19}, \cite{mt:06}, \cite {pod:08}, \cite{b:10}, \cite{lslm:14}, \cite{acmp:08} and \cite{zll:16}. Most of these versions are usually based on the following aspects: (A1) distributional properties to take into account positive skewness and an unimodal hazard rate function; (A2) a generalization of the linear form for the conditional mean (or median) dynamics; and (A3) a generalization of the properties of the time series; see \cite{b:10}.

A prominent generalization of ACD models is based on the Birnbaum-Saunders (BS) distribution; see \cite{b:10}. The BS-ACD model is constructed in terms of a conditional median duration and falls into all three categories (A1--A3). The BS-ACD model is built by taking into account the concept of conditional quantile estimation. Recent studies have shown that the BS-ACD model outperforms many other existing models in terms of model fit and forecasting performance; see \cite{sjla:19} . \cite{sl:17} proposed a family of ACD models based on the class of log-symmetric distributions, which contains the BS-ACD model as a special case. The class of log-symmetric distributions is a competitive and flexible alternative to deal with continuous, strictly positive, asymmetric, bimodal and/or light/heavy-tailed distributed data; see \cite{vp:16a}. Some examples of log-symmetric distributions are log-normal, log-Student-$t$, log-power-exponential, log-hyperbolic, log-slash, log-contaminated-normal, extended Birnbaum-Saunders and extended Birnbaum-Saunders-$t$. Log-symmetric ACD models cover all these log-symmetric distributions as special cases, that is, they cover highly competitive performance models in the literature.

Log-symmetric ACD models are written in terms of a conditional median duration rather than a conditional mean duration. The use of the median is more interesting because it is a better measure of central tendency than the mean, and the median-based approach provides additional protection against outliers. On the other hand, the log-symmetric family provides a wide range of asymmetric distributions with an unimodal hazard rate function. Therefore, features (C2) and (C3) and aspects (A1) and (A2) are treated by log-symmetric ACD models. The main advantage of log-symmetric ACD models is the robustness property of the median, that is, it is not affected by extremes or outliers. In terms of predictions, this implies that they will not be significantly affected by abnormal events; see \cite{sjla:19}.

In this work, we propose a new ACD model based on a class of log-symmetric distributions that is specified in terms of a time-varying conditional quantile duration instead of the traditionally employed conditional mean (or median) duration. The quantile approach allows us to capture the influences of conditioning variables on characteristics of the response distribution; see \cite{kx:06}. The proposed ACD model is based on quantile-based log-symmetric distributions introduced by \cite{sdlsf:22}. The quantile-based log-symmetric class of distributions has several desirable statistical properties that may make it preferable to alternative distributions. For example, the two parameters of the log-symmetric class are orthogonal and can be interpreted directly as quantile and skewness (or relative dispersion), which are, in the context of skew distributions, the most significant measures of location and shape, respectively. The flexibility provided by the log-symmetric family makes its corresponding ACD models an important area to be explored in the literature.  We illustrate the usefulness of the proposed ACD models by using a dataset on price durations of Apple Inc. stock. As a by-product, we derive a semi-parametric intraday value-at-risk (IVaR) model based on the proposed class of quantile-based log-symmetric ACD models. The proposed IVaR model is an adaptation of the one introduced by \cite{cht:07}. It provides the forecast of the quantile duration before the occurrence of the next price change and the forecast of the level of risk (IVaR). The IVaR measure is a very important tool for high frequency traders who have appeared with the development of financial markets and the consequent accessibility of high frequency data. IVaR is also a more adequate measure of risk when compared to the traditional VaR, as the risk must be determined within intervals that are smaller than daily time intervals. To the best of our knowledge, this is the first attempt to derive intradaily market risk models based on quantile ACD models.

The rest of this paper is organized as follows. In Section \ref{sec:02}, we briefly describe the quantile-based log-symmetric distributions. In Section \ref{sec:03}, we introduce the log-symmetric quantile ACD models. In this section, we also discuss inference and residual analysis. In Section \ref{sec:04}, we carry out Monte Carlo simulation studies for evaluating the performance of the maximum likelihood (ML) estimators as well as for assessing the empirical distribution of the residuals. In Section \ref{sec:05}, we apply the proposed ACD models to a real price duration data set and derive the corresponding IVaR model. Finally, in Section \ref{sec:06}, we provide some concluding remarks and discuss some future research work.


\section{Quantile-based log-symmetric distributions}\label{sec:02}

In this section, we briefly describe the family of quantile-based log-symmetric (QLS) distributions proposed by \cite{sdlsf:22}. These distributions will be useful for introducing the new QLS-ACD model. A random variable $X$ follows a QLS distribution, with quantile parameter $\Psi_q>0$ and power parameter $\phi>0$, if its probability density function (PDF) and cumulative distribution function (CDF) are given respectively by
\begin{equation}\label{fdp_quantil_log_simetrica}
f_X(x;\Psi_q, \phi) = \frac{\delta_{nc}}{\sqrt{\phi}\,x}g\left \{\frac{1}{\phi}
\left (\text{log}(x) - \text{log}(\Psi_q) + \sqrt{\phi}\,z_q\right)^2\right \},
\quad x>0, 
\end{equation}
and
\begin{equation}\label{fda_quantil_log_simetrica}
F_X(x;\Psi_q, \phi)=G\left (\frac{1}{\sqrt{\phi}}
\left (\text{log}(x) - \text{log}(\Psi_q) + \sqrt{\phi}\,z_q\right)\right ),
\quad x>0, 
\end{equation}
where $g(u)>0$, for $u>0$, is a kernel density generator function usually associated with an additional parameter $\vartheta$ (or extra parameter vector $\boldsymbol{\vartheta}$), $\delta_{nc}=1/\int_{-\infty}^\infty g(z^2){\rm d}z=1/\int_{0}^\infty u^{-1/2}g(u){\rm d}u$ is a normalization constant, $G(\omega) = \delta_{nc} \int_{-\infty}^{\omega} g(z^2) \mathrm{d}z$ with  $\omega \in \mathbb{R}$, and ${z_{q}}=G^{-1}(q)$. Let us denote $X \sim \mathrm{QLS}(\Psi_q,\phi,g)$. Note that the parameter $\Psi_q$ is the $q$-th quantile of $X$ and the parameter $\phi$ denotes the skewness (or the relative dispersion); see \cite{dsvfp:23}. Table \ref{Tabela_gerador_de_densidade} presents some special cases of the quantile-based log-symmetric family of distributions.

\begin{table}[h!]
	\small
	\centering
	\renewcommand{\arraystretch}{1.1}
	\caption{Density generator $g(u)$ for some log-symmetric distributions.}\label{Tabela_gerador_de_densidade}
	{
		\begin{tabular}{ll}
			\toprule
			Distribution                    &$g(u)$                                                                                   \\
			\midrule
		Log-normal($\lambda,\phi$) & $\exp\left( -\frac{1}{2}u\right)$\\
		Log-Student-$t$($\lambda,\phi,\vartheta$) & $\left(1+\frac{u}{\vartheta} \right)^{-\frac{\vartheta+1}{2}}$, $\vartheta>0$\\
		Log-power-exponential($\lambda,\phi,\vartheta$) & $\exp\left( -\frac{1}{2}u^{\frac{1}{1+\vartheta}}\right)$, $-1<{\vartheta}\leq{1}$\\
		Log-hyperbolic($\lambda,\phi,\vartheta$) & $\exp(-\vartheta\sqrt{1+u})$, $\vartheta>0$\\
		Log-slash($\lambda,\phi,\vartheta$)& $\textrm{IGF}\left(\vartheta+\frac{1}{2} ,\frac{u}{2}\right)$, $\vartheta>0$\\
		Log-contaminated-normal($\lambda,\phi,{\bm\vartheta}=(\vartheta_1,\vartheta_2)^\top$)& $\sqrt{\vartheta_2}\exp\left(-\frac{1}{2}\vartheta_2 u\right)+\frac{(1-\vartheta_1)}{\vartheta_1},
		\exp\left(-\frac{1}{2} u\right)$, $0<\vartheta_1,\vartheta_2<1$\\
		Extended Birnbaum-Saunders($\lambda,\phi,\vartheta$)& $\cosh(u^{1/2})\exp\left(-\frac{2}{\vartheta^2}\sinh^2(u^{1/2}) \right)$, $\vartheta>0$\\
		Extended Birnbaum-Saunders-$t$($\lambda,\phi,{\bm\vartheta}=(\vartheta_{1},\vartheta_{2})^{\top}$)&
		$\cosh(u^{1/2})\left(\vartheta_{2}\vartheta_{1}^2+4 \sinh^2(u^{1/2})\right)^{-\frac{\vartheta_{2}+1}{2}}$, $\vartheta_{1},\vartheta_{2}>0$\\
			\bottomrule
		\end{tabular}
	}
\end{table}


\section{QLS-ACD Model}\label{sec:03}

\subsection{Formulation} \label{sec:03.01}

Let $\{\tau_1, \dots, \tau_n\}$ be a sequence of successive times when market events, or trades, occur. The duration or the elapsed time between two transactions, $\tau_t$ and $\tau_{t-1}$, is defined by $x_t=\tau_t-\tau_{t-1}$ for $t = 1, \dots, n$ . The proposed QLS-ACD model is specified in terms of a time-varying conditional quantile, $\Psi_{q,t} = F_X^{-1}(q;\Psi_{q,t}, \phi|\mathcal{F}_{t-1})$, given $q\in(0,1)$, where, $F_{X_t}^{-1}$ is the quantile function of the QLS distribution and $\mathcal{F}_{t-1}$ is the information set which includes all available information up to $\tau_{t-1}$. Thus, the conditional PDF of the QLS-ACD model is given by
\begin{equation}\label{fdp_quantil_log_simetrica_ACD}
f_{X_t|\mathcal{F}_{t-1}}(x_t;\Psi_{q,t}, \phi) = \frac{\delta_{cn}}{\sqrt{\phi}\,x_t}g\left( \frac{1}{\phi}
\left (\text{log}(x_t) - \text{log}(\Psi_{q,t}) + \sqrt{\phi}\,z_q\right)^2\right ),
\quad x_t>0, 
\end{equation}
where $\Psi_{q,t}$, $\phi$ and $g(\cdot)$ are the conditional quantile of $X_t$, the skewness (or relative dispersion) parameter and the kernel density generating function, respectively. Then, we define the dynamics of the time-varying conditional quantile of the QLS-ACD model as follows:
\begin{eqnarray}\label{dynamic_conditional_quantile}
\eta_t=h(\Psi_{q, t}) = \omega + \sum_{j=1}^{r}\alpha_j \text{log}(\Psi_{q, t-j}) + \sum_{j=1}^{s}\beta_j \left(\frac{x_{t-j}}{\Psi_{q, t-j}}\right),\quad t = 1, \dots, n,
\end{eqnarray}
where $\omega$, $\boldsymbol{\alpha} = (\alpha_1,\ldots,\alpha_r)^\top \in \mathbb{R}^r$ and $\boldsymbol{\beta} = (\beta_1,\ldots,\beta_s)^\top \in \mathbb{R}^s$ are ACD parameters, which leads to the notation QLS-ACD($r, s, q, g$). $h: \mathbb{R} \rightarrow \mathbb{R}^{+}$  is a continuously twice differentiable strictly increasing link function with inverse given by $h^{-1}: \mathbb{R}^{+}  \rightarrow \mathbb{R}$, which is twice continuously differentiable as well. The ACD model in \eqref{fdp_quantil_log_simetrica_ACD} includes as special case the log-symmetric model ACD$(r, s)$ proposed by \cite{sl:17} when $q=0.5$ with a logarithmic link for $h$. In relation to the link function $h$ in  \eqref{dynamic_conditional_quantile}, a common choice is the logarithm (log) link as it does not impose non-negativity restrictions on the parameters; see \cite{bg:00}.

\subsection{Estimation} \label{sec:03.02}

The estimates of the parameters of the QLS-ACD($r,s,q$) model can be obtained by using the ML method as in \cite{b:10}. Let $\boldsymbol{\theta} = (\phi,\omega,\boldsymbol{\alpha}^\top,\boldsymbol{\beta}^\top)^\top$ denote the parameter vector. Then, the likelihood function is given by
	\begin{equation}
		\bm{L}(\boldsymbol{\theta}) = \prod_{t = 1}^{n} f_{X_t|\mathcal{F}_{t-1}}(x_t;\Psi_{q, t}, \phi),\quad x_t > 0, X_t|\mathcal{F}_{t-1} \sim \mathrm{QLS}(\Psi_{q, t},\phi,g),
	\end{equation}
	which implies that the log-likelihood function (without the constant term) can be expressed as
	\begin{equation}\label{eq:loglik}
		\bm{\ell}(\boldsymbol{\theta}) = \sum_{t = 1}^{n} \log(g(z_t^2)) - \frac{n}{2}\log(\phi),
	\end{equation}
	where $z_t = \left[ \log(x_t)-\log(\Psi_{q, t}) + \sqrt{\phi}{z_{q}} \right]/\sqrt{\phi} $, $t = 1, \ldots,n$, and $\Psi_{q, t}$ is as given in \eqref{dynamic_conditional_quantile}.
To obtain an estimate of $\boldsymbol{\theta} = (\phi,\omega,\boldsymbol{\alpha}^\top,\boldsymbol{\beta}^\top)^\top$, we equate the score vector containing the first-order partial derivatives of $\bm{\ell}(\boldsymbol{\theta})$, $\dot{\bm{\ell}}_{\boldsymbol{\theta}} = (\dot{\bm \ell}_{\phi},\dot{\bm \ell}_{\omega},\dot{\bm \ell}_{\alpha_l}^\top,\dot{\bm \ell}_{\beta_m}^\top)^\top$ say, to zero vector, providing the likelihood equations. Here,
			\begin{eqnarray*}
				\dot{\bm \ell}_{\phi} &=& \frac{1}{2\phi } \sum _{t=1}^n {z_t(z_t-z_q)v\left(z_t\right)} -\frac{n}{2 \phi }, \quad
				\dot{\bm \ell}_{\omega} = \frac{1}{\sqrt{\phi}}\sum _{t=1}^n \frac{\partial \Psi _{q,t} }{\partial \omega}   \frac{z_t}{\Psi _{q,t}} v\left(z_t\right),\nonumber \\
				\dot{\bm \ell}_{\alpha_l} &=& \frac{1}{\sqrt{\phi}}\sum _{t=1}^n \frac{\partial \Psi _{q,t} }{\partial \alpha_l}   \frac{z_t}{\Psi _{q,t}} v\left(z_t\right), l=1,\ldots,r, \quad
				\dot{\bm \ell}_{\beta_m} = \frac{1}{\sqrt{\phi}}\sum _{t=1}^n \frac{\partial \Psi _{q,t} }{\partial \beta_m}   \frac{z_t}{\Psi _{q,t}} v\left(z_t\right), m=1,\ldots,s, \nonumber \\
			\end{eqnarray*}
			where
			$v(u) = -2g'{(u^2)}/g(u^2)$, with $g'(u) = \mathrm{d} g(u)/\mathrm{d}u$, are weights associated with the $g(\cdot)$ function, and
\begin{eqnarray*}
{\partial \Psi _{q,t}\over \partial \omega}
&=&
\left(
\sum_{j=1}^r {\alpha_j\over\Psi _{q,t-j}}
{\partial \Psi _{q,t-j}\over \partial \omega}
-
\sum_{j=1}^s {\beta_j x_{t-j}\over \Psi^2_{q,t-j}}
{\partial \Psi _{q,t-j}\over \partial \omega}
\right) \Psi _{q,t},
\\[0,2cm]
{\partial \Psi _{q,t}\over \partial \alpha_l}
&=&
\left(
\alpha_l \log( \Psi _{q,t-l}) +
\sum_{j=1}^r {\alpha_j\over\Psi _{q,t-j}}
{\partial \Psi _{q,t-j}\over \partial \alpha_l}
-
\sum_{j=1}^s {\beta_j x_{t-j}\over \Psi^2_{q,t-j}}
{\partial \Psi _{q,t-j}\over \partial \alpha_l}
\right) \Psi _{q,t},
\quad l=1,\ldots,r,
\\[0,2cm]
{\partial \Psi _{q,t}\over \partial \beta_m}
&=&
\left(
\sum_{j=1}^r {\alpha_j\over\Psi _{q,t-j}}
{\partial \Psi _{q,t-j}\over \partial \beta_m}
+
\beta_m {x_{t-m}\over \Psi _{q,t-m}}
-
\sum_{j=1}^s {\beta_j x_{t-j}\over \Psi^2_{q,t-j}}
{\partial \Psi _{q,t-j}\over \partial \beta_m}
\right) \Psi _{q,t},
\quad m=1,\ldots,s.
\end{eqnarray*}
We solve the likelihood equations by using the Broyden-Fletcher-Goldfarb-Shanno (BFGS) algorithm. Starting values are obtained from the \texttt{R} function \texttt{acdFit}, which is associated with the \texttt{ACDm} package; see \cite{be:22}. The extra parameter ${\vartheta}$ (or extra parameter vector ${\bm\vartheta}$) is estimated by the profile log-likelihood method, which encompasses three steps:
\begin{itemize}
 \item[1)] Let $\vartheta^{(k)}=k$ and for each $k=k_L,\ldots,k_U$, where $k_L$ and $k_U$ are lower and upper predefined limits, respectively, compute the ML estimate of $\boldsymbol{\theta}$.
 \item[2)] Compute the value of the log-likelihood function with the estimates obtained in Step 1;
 \item[3)] Set the final estimate of $\vartheta$ from $\{\vartheta^{(k_L)},\ldots,\vartheta^{(k_U)}\}$ as the one that maximizes the log-likelihood function and the associated estimate of $\bm\theta$ is then the final one.
\end{itemize}

Inference on $\boldsymbol{\theta}$ can be based on the asymptotic distribution of the ML estimator $\boldsymbol{\widehat{\theta}}$, which under regularity conditions (see Assumptions 2.1-2.5 of \citealt{Andersen:70}) and sufficiently large $n$, is distributed as multivariate normal, that is,
\begin{equation}
		\sqrt{n}(\boldsymbol{\widehat{\theta} - \theta}) \build{\longrightarrow}{\cal D}{} \mathrm{N}_{2+r+s}(\boldsymbol{0}, {\cal I}(\boldsymbol{\theta})^{-1}) \nonumber
\end{equation}
as $n \to \infty$, where $\build{\longrightarrow}{\cal D}{}$ denotes ``convergence in distribution'' and ${\cal I}(\boldsymbol{\theta})$ denotes the expected Fisher information matrix. In practice, this matrix can be approximated by its observed version obtained from the Hessian matrix, which in this work is obtained numerically; the \texttt{R} function \texttt{maxBFGS}, which is associated with the \texttt{maxLik} package, is used to compute numerically the elements of the Hessian matrix.

\subsection{Residual analysis} \label{sec:03.03}

Residual analysis plays a key role in validating any statistical model. Such analysis allows detecting the existence of possible discrepant observations (\textit{outliers}) and their effects on the inference or on the adjustment results. In particular, 
we consider the generalized Cox-Snell residual (GCS), which is generally used to assess the overall goodness-of-fit of the model and is widely used for data with asymmetric responses. The GCS residuals are defined as
\begin{equation}\label{RES_COX_SNEL_GEN_RBS}
r_{t}^{GCS}=-\text{log}(\widehat{S}_{X_t|\mathcal{F}_{t-1}}(x_t;\widehat{\Psi}_{q,t},\widehat{\phi})),
\end{equation}
where $\widehat{S}_{X_t|\mathcal{F}_{t-1}}(x_t;\widehat{\Psi}_{q,t},\widehat{\phi})$ is the estimate of the survival function. According to \cite{b:10}, the GCS residuals follow a standard exponential distribution, EXP(1), when the model has a correct specification; see \cite{dsvfp:23}. Then,  quantile-quantile (QQ) plots with simulated envelope of $r_{t}^{GCS}$ can be used to assess model fit.


\section{Monte Carlo Simulation}\label{sec:04}

In this section, we present the results of a Monte Carlo simulation study for the QLS-ACD($r, s, q$) model to assess the performance of the ML estimators and the empirical distribution of the GCS residuals. We consider the QLS-ACD$(1,1,q)$ model based on the log-normal distribution (similar results were obtained for other quantile-based log-symmetric models and are not presented for the sake of conciseness). In practice, a higher order for ACD models do not improve the model fit considerably. Hence, the conditional quantile dynamics is as follows:
\begin{eqnarray*}
 \log(\Psi_{q,t})= \omega + \alpha \log(\Psi_{q,t-1}) + \beta\frac{x_{t-1}}{\Psi_{q,t-1}}.
 \end{eqnarray*}
We consider the following true choices of the parameters: $(\phi,\omega,\alpha,\beta) = (0.25,0.20,0.70,0.10)$. Moreover, different values of quantile and sample sizes are considered: $q=\{0.05,0.25,0.50,0.75,$ $0.95\}$ and $n \in \left\lbrace 200,1000,2000\right\rbrace$. We consider $N=500$ Monte Carlo replications for each sample size. To asses the performance of the ML estimators, we compute the corresponding relative biases (RBs) and root mean squared errors (RMSEs). Moreover, to assess the empirical distribution of the GCS residuals, we monitor the following descriptive statistics: mean, median, standard deviation, coefficient of skewness and coefficient of excess  kurtosis. The steps involved in the study are presented in Algorithm~\ref{alg:simulation}.

\begin{algorithm}[!ht]
\floatname{algorithm}{Algorithm}
\caption{Steps involved in the simulation study.}\label{alg:simulation}
\begin{algorithmic}[1]
\State Generate 500 samples based on the quantile-based log-normal ACD model.
\State Obtain the ML estimates of the model parameters for each sample.
\State Compute the empirical RB and MSE of the estimators:
\begin{eqnarray*}
 \widehat{\textrm{RB}}(\widehat{\theta}) =  \left|\frac{\frac{1}{N} \sum_{i = 1}^{N} \widehat{\theta}^{(i)} - \theta}{\theta}\right| , \quad
\widehat{\mathrm{RMSE}}(\widehat{\theta}) = {\sqrt{\frac{1}{N} \sum_{i = 1}^{N} (\widehat{\theta}^{(i)} - \theta)^2}},
\end{eqnarray*}
where $\theta$ and $\widehat{\theta}^{(i)}$ are the true parameter value and its $i$-th ML estimate, and $N$ is the number of Monte Carlo replications.
\State For each Monte Carlo run, calculate the GCS residuals according to \eqref{RES_COX_SNEL_GEN_RBS}, and the respective descriptive statistics: mean, median, standard deviation, coefficient of skewness and coefficient of excess kurtosis.
\State Calculate the means of the descriptive statistics obtained in Step 4 based on 500 runs.
\end{algorithmic}
\end{algorithm}

\begin{figure}[H]
\vspace{-0.25cm}
\centering
{\includegraphics[height=4cm,width=4cm]{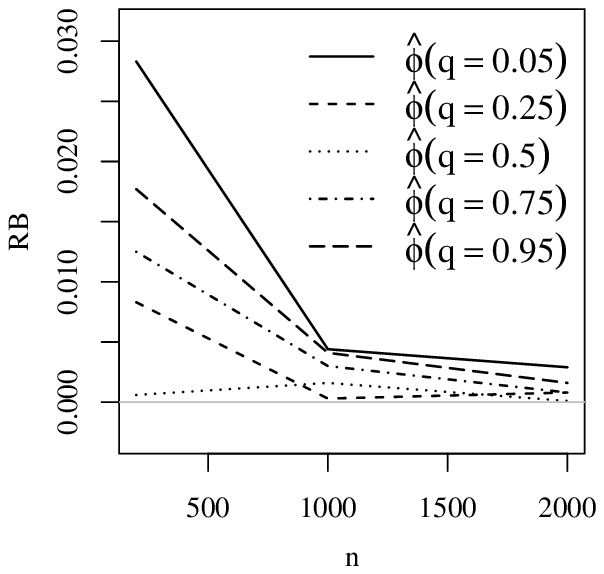}}\hspace{-0.25cm}
{\includegraphics[height=4cm,width=4cm]{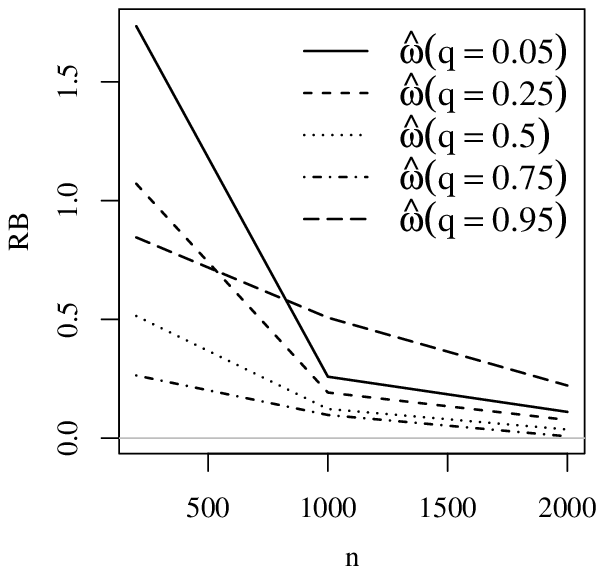}}\hspace{-0.25cm}
{\includegraphics[height=4cm,width=4cm]{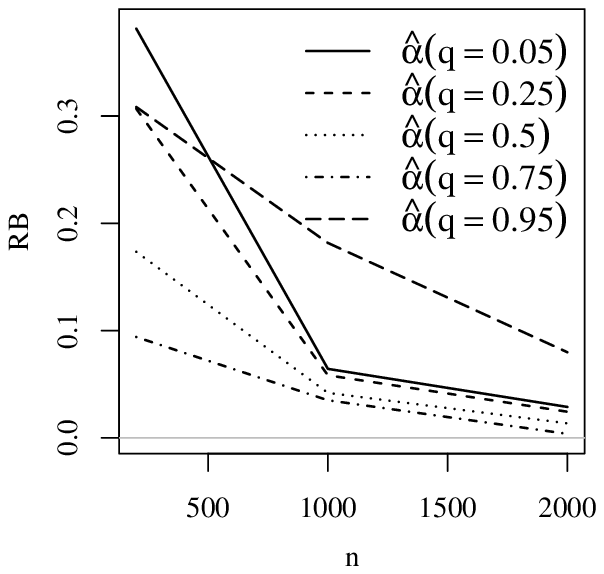}}\hspace{-0.25cm}
{\includegraphics[height=4cm,width=4cm]{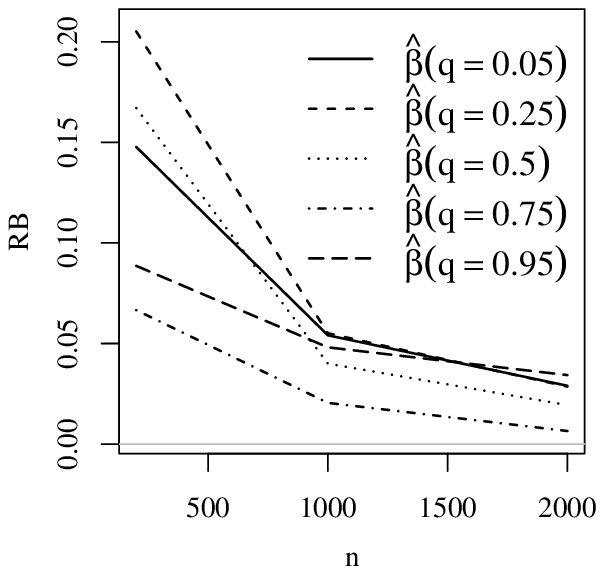}}\hspace{-0.25cm}\\
{\includegraphics[height=4cm,width=4cm]{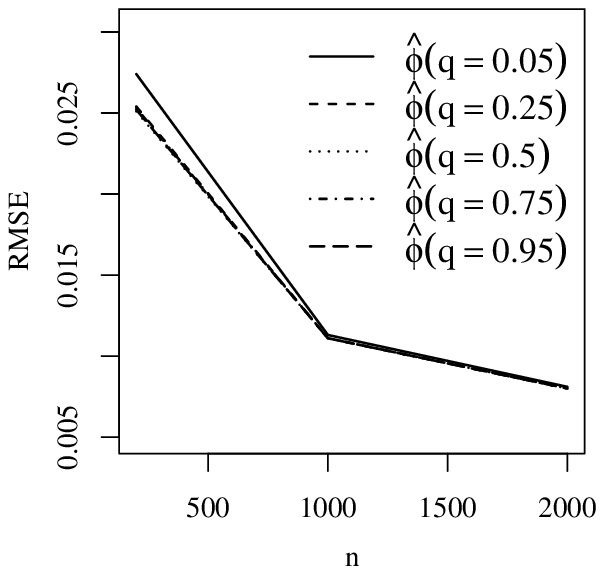}}\hspace{-0.25cm}
{\includegraphics[height=4cm,width=4cm]{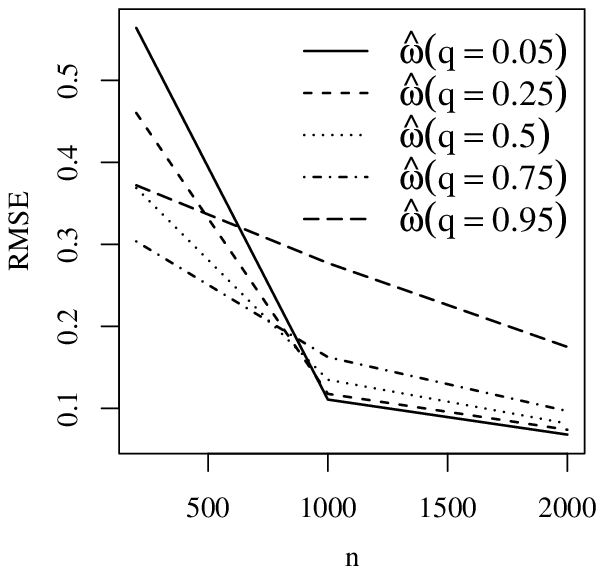}}\hspace{-0.25cm}
{\includegraphics[height=4cm,width=4cm]{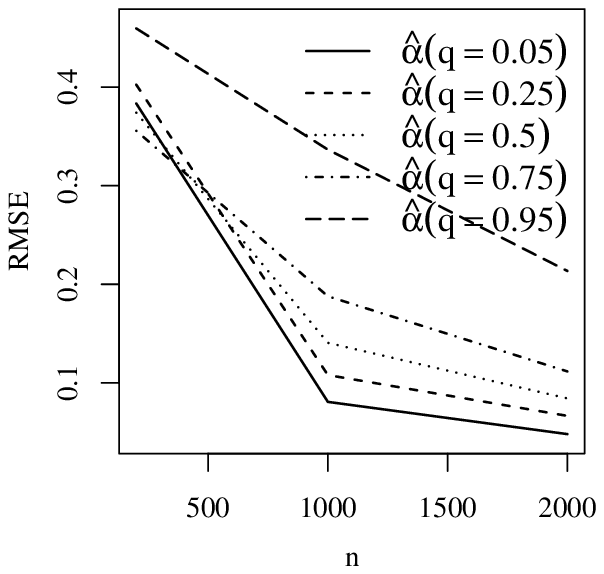}}\hspace{-0.25cm}
{\includegraphics[height=4cm,width=4cm]{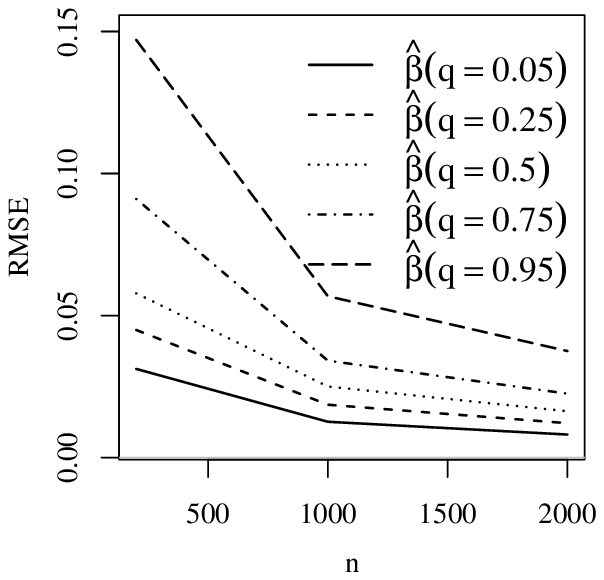}}\hspace{-0.25cm}
\vspace{-0.2cm}
\caption{Monte Carlo simulation results for the quantile-based log-normal ACD model.}
\label{fig_normal_mc}
\end{figure}

The ML estimation results are presented in Figure~\ref{fig_normal_mc} wherein the empirical RB and RMSE are both displayed. Since the ML estimators are consistent, we expect the RB and RMSE to approach zero as $n$ increases From Figure~\ref{fig_normal_mc}, we indeed observe the RB and RMSE to decrease with an increase in sample size.
%

\begin{figure}[!ht]
\vspace{-0.25cm}
\centering
{\includegraphics[height=4.5cm,width=4.5cm]{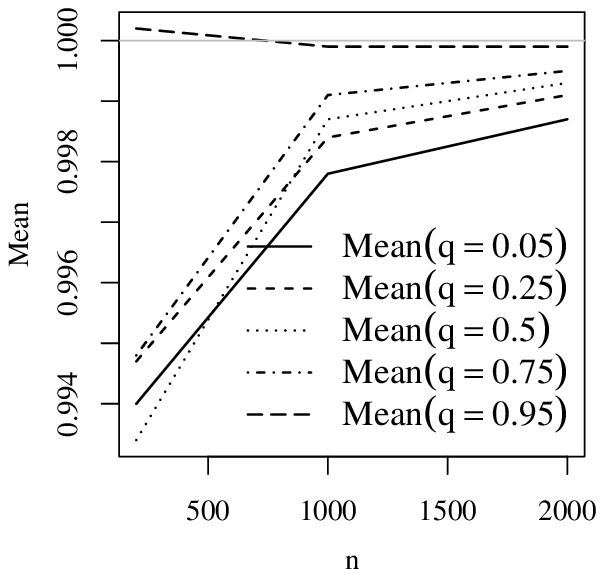}}\hspace{-0.25cm}
{\includegraphics[height=4.5cm,width=4.5cm]{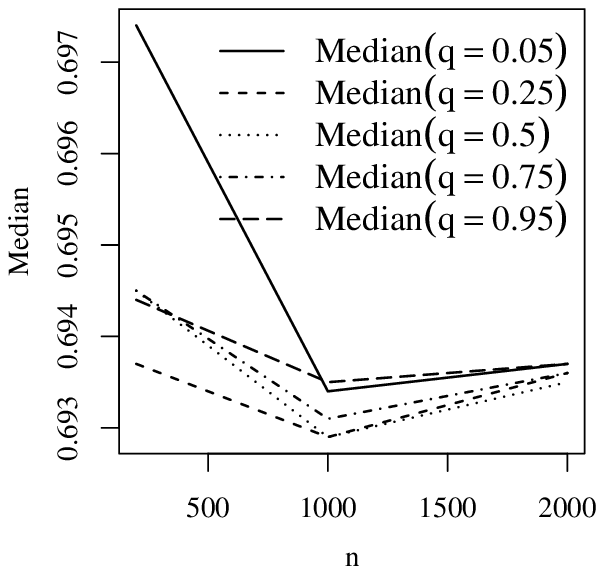}}\hspace{-0.25cm}
{\includegraphics[height=4.5cm,width=4.5cm]{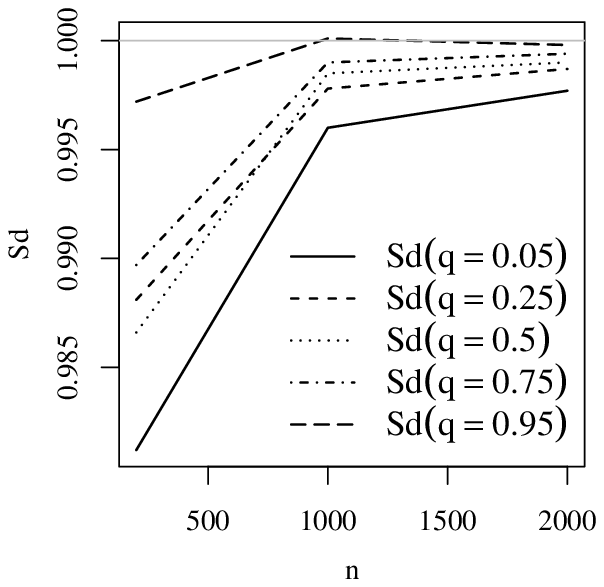}}\hspace{-0.25cm}
{\includegraphics[height=4.5cm,width=4.5cm]{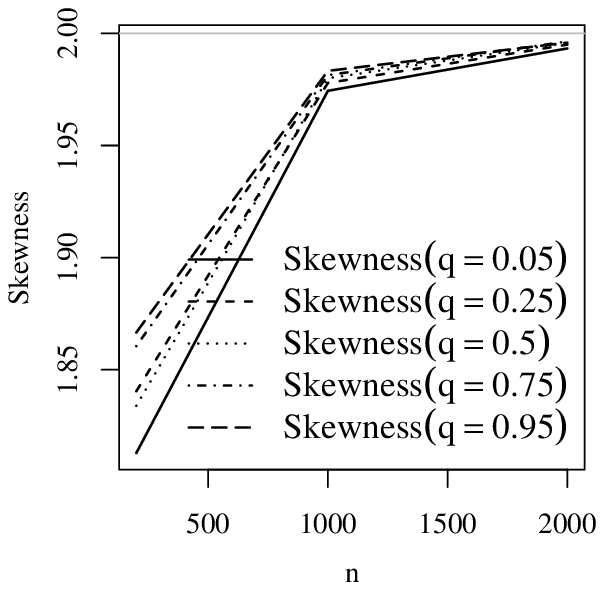}}\hspace{-0.25cm}
{\includegraphics[height=4.5cm,width=4.5cm]{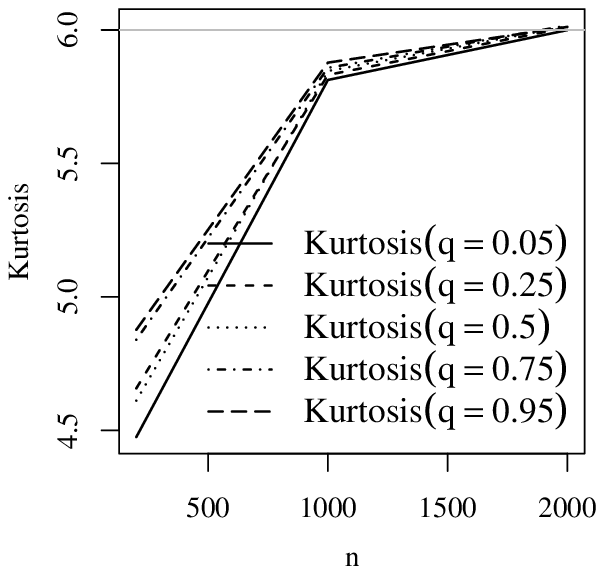}}
\vspace{-0.2cm}
\caption{Monte Carlo simulation results of the GCS residuals for the quantile-based log-normal ACD model.}
\label{fig_normal_MC_GCS}
\end{figure}

Figure \ref{fig_normal_MC_GCS} shows the simulation results for the $r_t^\textrm{GCS}$ residuals. As the reference distribution for $r_t^\textrm{GCS}$ is EXP(1), the values of mean, median, standard deviation, coefficient of skewness and coefficient of excess kurtosis are expected to be 1, 0.69, 1, 2 and 6, respectively. From, Figure \ref{fig_normal_MC_GCS}, we observe that, in general, the considered residuals conform well with the reference distribution, and we can therefore use the GCS residuals to assess the fit of the proposed model.

\newpage

\section{Application to intraday value-at-risk}\label{sec:05}

In this section, price durations of Apple Inc. stock on 15th August 2023 (\url{www.dukascopy.com}) are used to illustrate the proposed QLS-ACD models. Initially, we present how to compute Apple price durations and present an exploratory data analysis to verify if the proposed models are a reasonable assumption. We then fit the models to investigate the best model among the class of QLS-ACD models. Finally, we propose an algorithm for IVaR forecasting based on the best model.

\subsection{Apple price durations and exploratory data analysis}\label{sec:05.1}

Let $(\tau_t,p_{\tau_t})$, $t=1,\ldots,n$, be a market point process, where $\tau_t$ denotes the time for which prices ($p_{\tau_t}$) have increased or decreased by at least $\kappa$, and $p_{\tau_t}=(a_{\tau_t}+b_{\tau_t})/2$ denotes the mid-point of the bid ($b_{\tau_t}$) and ask ($a_{\tau_t}$) prices. The price duration between two price changes is given by $x_t=\tau_t-\tau_{t-1}$ (see Section \ref{sec:03.01}) and the price changes returns are given by $r_{\tau_t}=\log(p_{\tau_t})-\log(p_{\tau_{t-1}})$. Hereafter, we consider seasonally adjusted price durations, $\{\tilde{x}_t\}_{t=1}^{n}$, where
 $\tilde{x}_t=\frac{{x}_t}{\widehat{\varpi}_t}$, $t=1,\ldots,n$, with ${x}_t$ and $\widehat{\varpi}_t$ denoting the raw price durations and the intraday seasonality, respectively. For the Apple data, using $\kappa=0.01$, we computed the corresponding price durations and removed their intraday seasonality by using the \texttt{R} functions \texttt{computeDurations} and \texttt{diurnalAdj}, which are associated with the \texttt{ACDm} package; see \cite{be:22}.

Descriptive statistics for both raw and adjusted price durations are provided in Table~\ref{tab:descp_ex}. From these statistics, we observe that mean is greater than the median for both series, which indicates the presence of positive skewness. Moreover, the coefficient of kurtosis suggests occurrence of heavy tails, and the coefficient of skewness confirms the presence of positive skewness.

\begin{table}[!ht]
\centering
\caption{{Summary statistics for the Apple data.}}\label{tab:descp_ex}
\begin{tabular}{lcccccccc}
\hline
BASF-SE data                       &    Plain   & &   Adjusted   \\
\hline
$n$                                &   6689     & &  6689    \\
Minimum                            &   1        & &  0.251    \\
10th percentile                    &  1         & &  0.283         \\
Mean                               &  2.96      & &  1.019      \\
50th percentile (median)           &   2        & & 0.666       \\
90th percentile                    &  6         & &  2.155     \\
Maximum                            &  3.187     & & 0.986       \\
Standard deviation                 & 9.877      & & 1.579       \\
Coefficient of variation           &  107.676\% & & 96.757\%        \\
Coefficient of skewness            & 3.165      & & 2.91     \\
Coefficient of excess kurtosis     & 14.448     & & 13.357     \\
\hline
\end{tabular}
\end{table}

\begin{figure}[!ht]
\centering
\subfigure[]{\includegraphics[height=5cm,width=5cm]{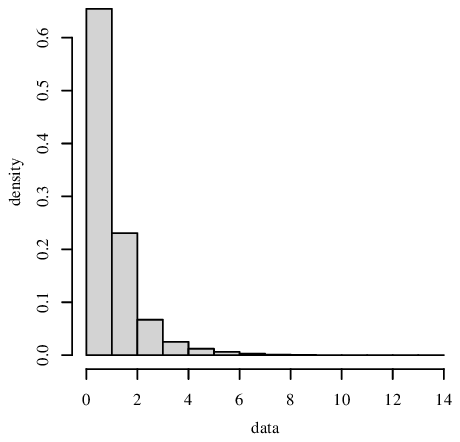}}
\subfigure[]{\includegraphics[height=5cm,width=5cm]{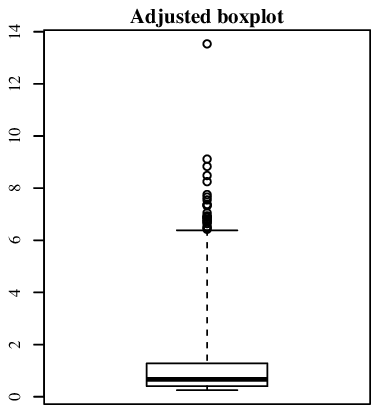}}
\subfigure[]{\includegraphics[height=5cm,width=5cm]{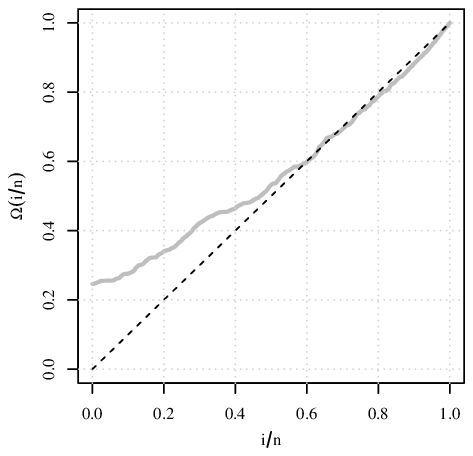}}
 \caption{\small {Histogram, adjusted boxplot and TTT plot for the Apple data.}}
\label{fig:histttt_ex}
\end{figure}

Figure \ref{fig:histttt_ex} presents the histogram, adjusted boxplot and total time on test (TTT) plot for the Apple data. Figure \ref{fig:histttt_ex}(a) confirms the positive skewness suggested in the descriptive statistics. Moreover, the adjusted boxplot \citep{hv:08} displayed in Fig.~\ref{fig:histttt_ex}(b) indicates potential outliers. Finally, the TTT plot indicates an unimodal hazard rate for the Apple data set. Therefore, the descriptive statistics and the plots of the histogram, adjusted boxplot and TTT all reveal that the proposed QLS-ACD models seem to be a reasonable assumption for these data, owing to their ability to account for asymmetry, heavy tails and an unimodal hazard rate.

\subsection{Estimation results for the QLS-ACD models}\label{sec:05.2}

\begin{figure}[!ht]
\centering
\subfigure[Log-normal]{\includegraphics[height=4.0cm,width=4.0cm]{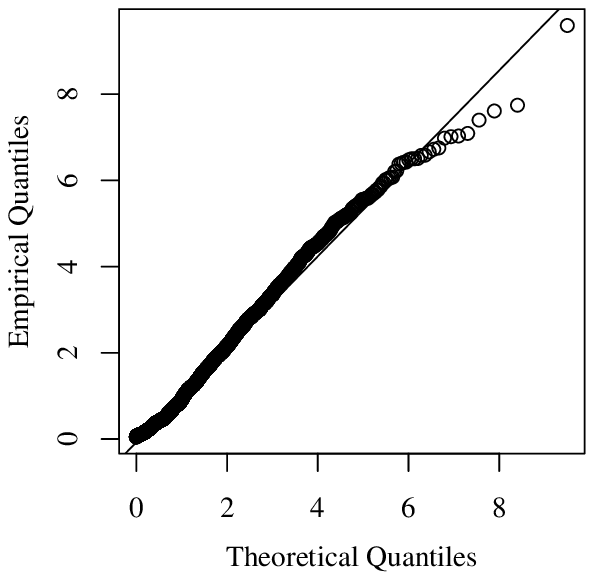}}
\subfigure[Log-Student-$t$]{\includegraphics[height=4.0cm,width=4.0cm]{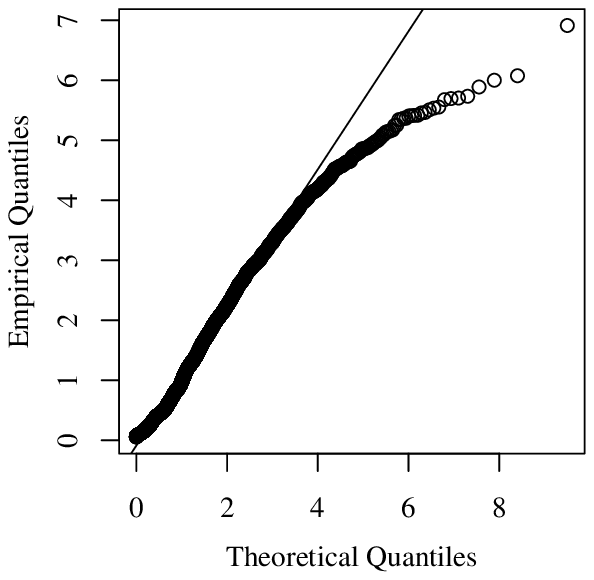}}
\subfigure[Log-power-exponential]{\includegraphics[height=4.0cm,width=4.0cm]{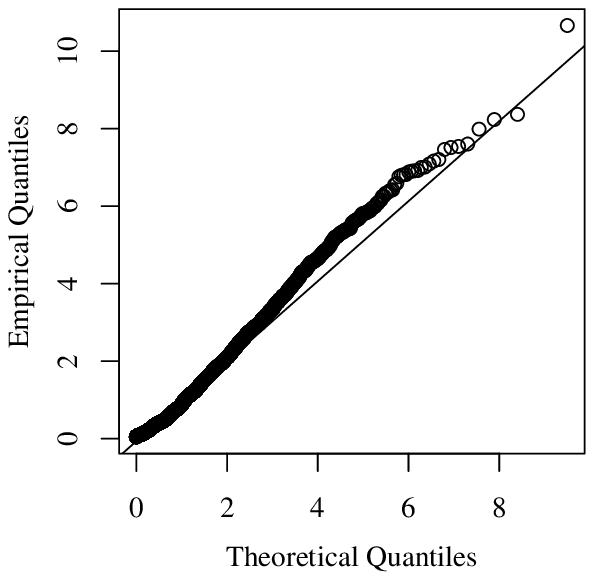}}
\subfigure[Log-hyperbolic]{\includegraphics[height=4.0cm,width=4.0cm]{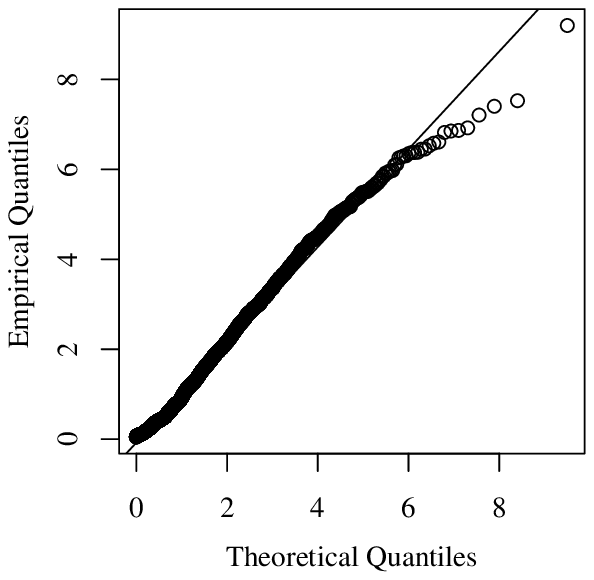}}\\
\subfigure[Log-slash]{\includegraphics[height=4.0cm,width=4.0cm]{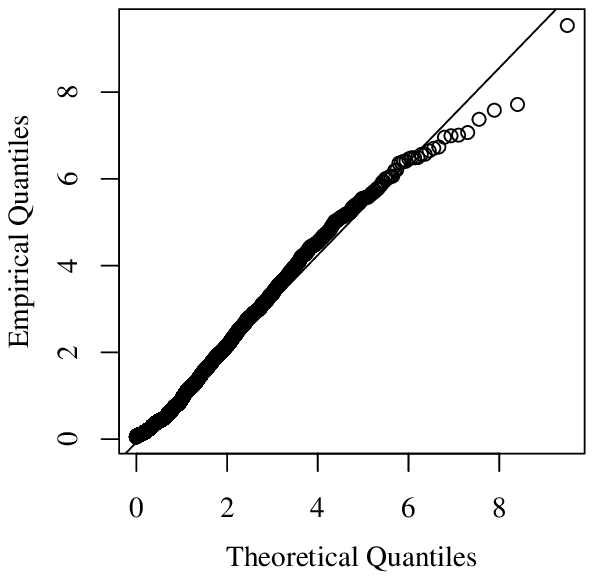}}
\subfigure[Log-contaminated-normal]{\includegraphics[height=4.0cm,width=4.0cm]{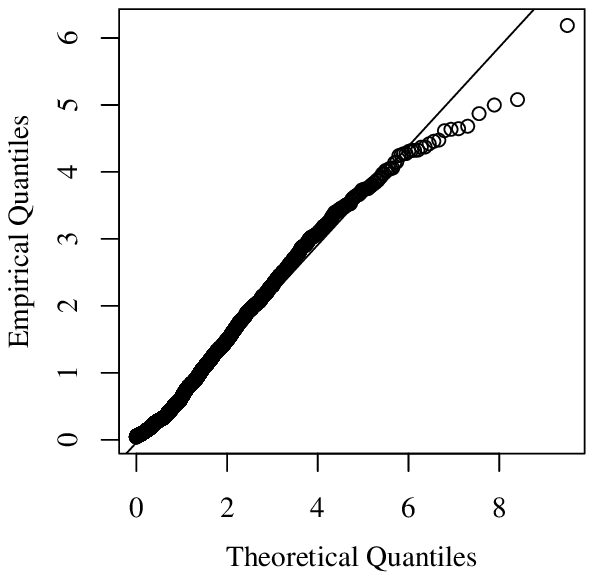}}
\subfigure[Extended Birnbaum-Saunders]{\includegraphics[height=4.0cm,width=4.0cm]{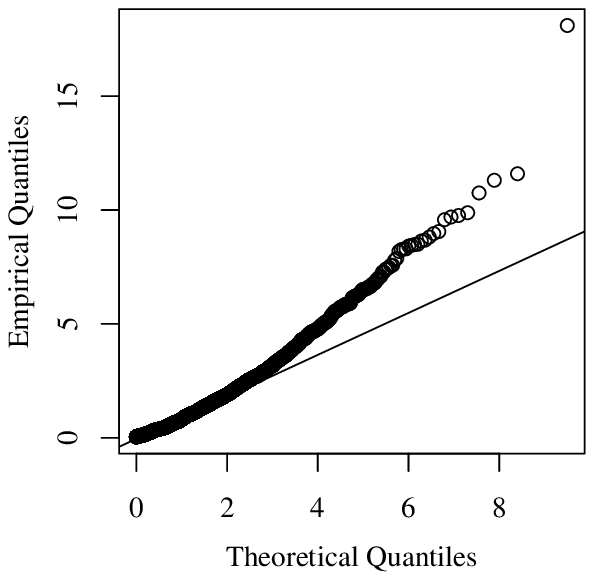}}
\subfigure[Extended Birnbaum-Saunders-$t$]{\includegraphics[height=4.0cm,width=4.0cm]{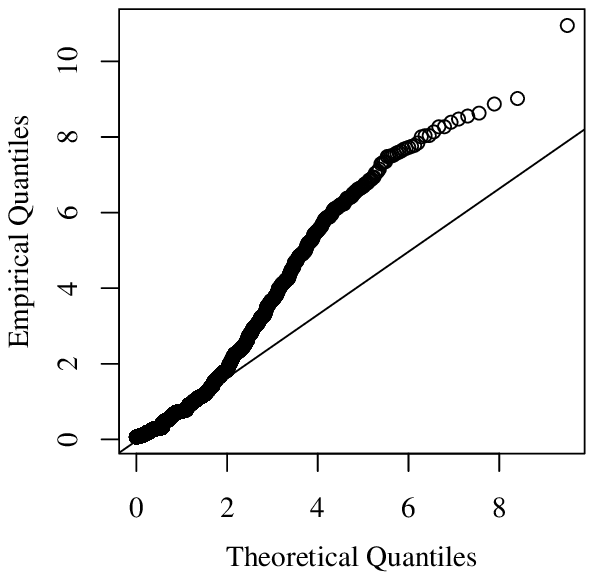}}

 \caption{\small {QQ plot and its envelope for the generalized Cox-Snell residuals in the indicated model for the Apple data ($q=0.025$).}}
\label{fig:qqplots_q0025}
\end{figure}

\begin{figure}[!ht]
\centering
\subfigure[Log-normal]{\includegraphics[height=4.0cm,width=4.0cm]{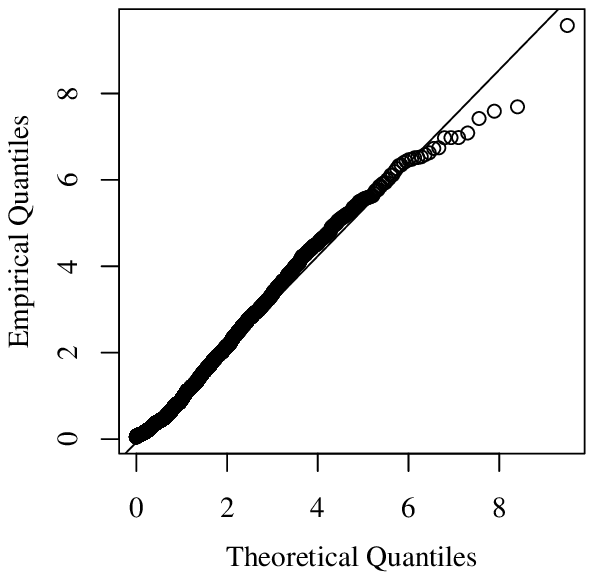}}
\subfigure[Log-Student-$t$]{\includegraphics[height=4.0cm,width=4.0cm]{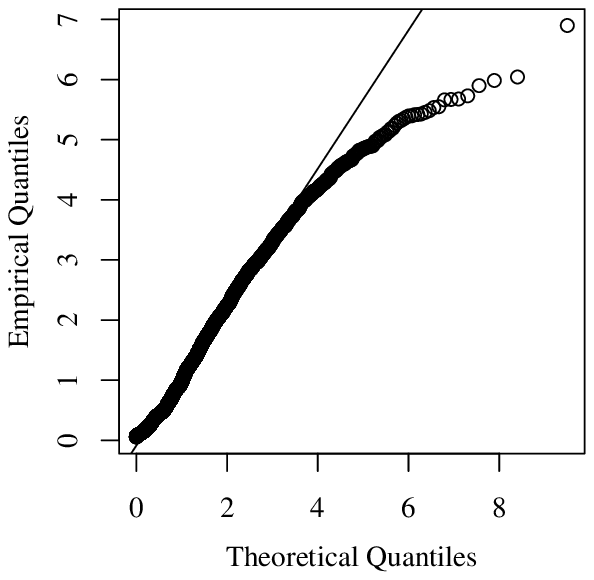}}
\subfigure[Log-power-exponential]{\includegraphics[height=4.0cm,width=4.0cm]{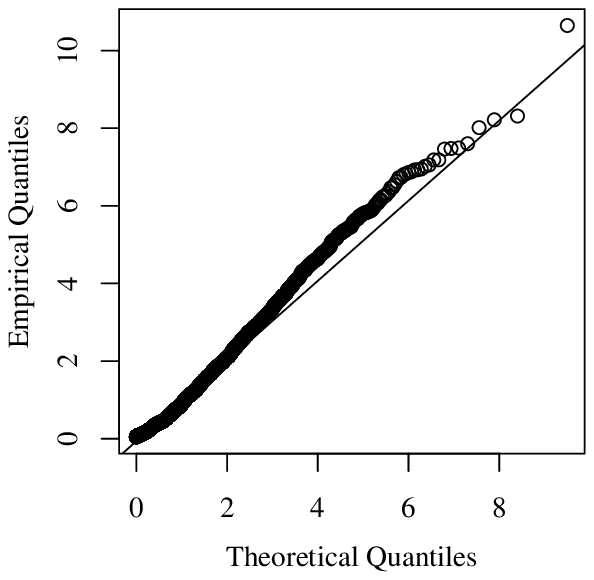}}
\subfigure[Log-hyperbolic]{\includegraphics[height=4.0cm,width=4.0cm]{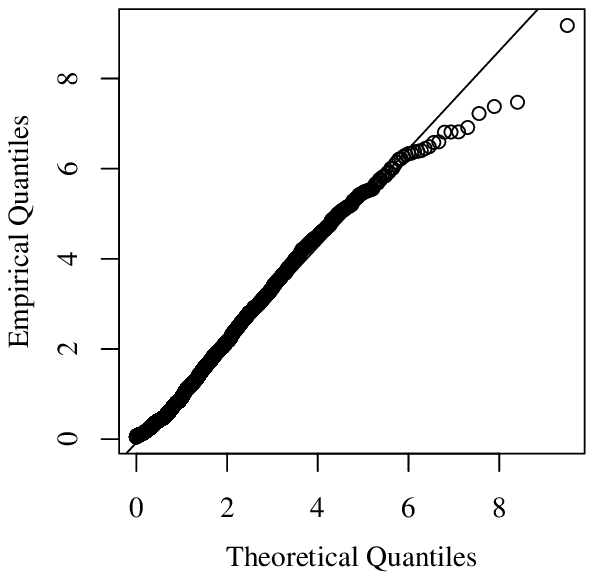}}\\
\subfigure[Log-slash]{\includegraphics[height=4.0cm,width=4.0cm]{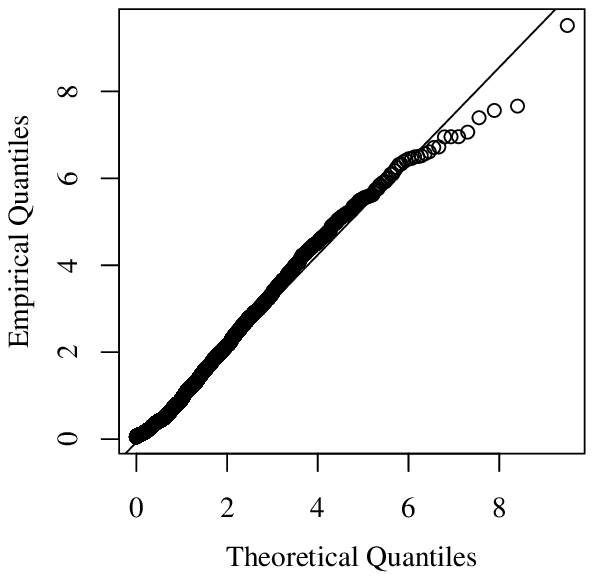}}
\subfigure[Log-contaminated-normal]{\includegraphics[height=4.0cm,width=4.0cm]{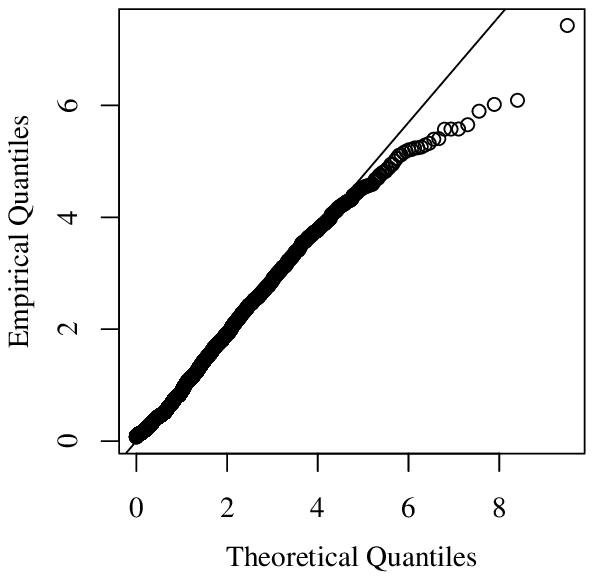}}
\subfigure[Extended Birnbaum-Saunders]{\includegraphics[height=4.0cm,width=4.0cm]{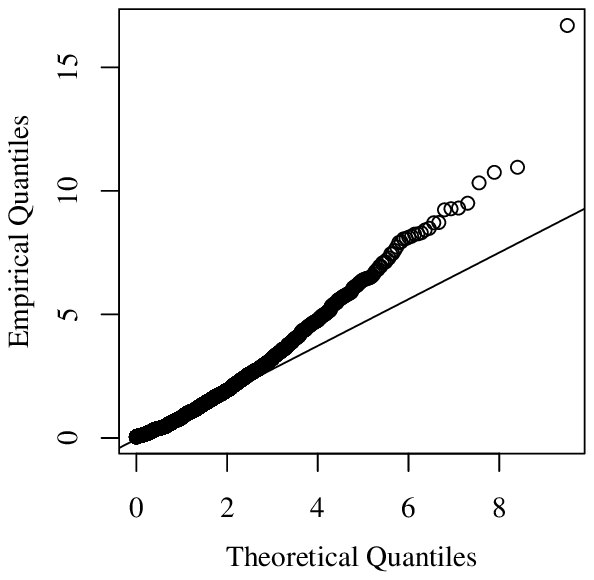}}
\subfigure[Extended Birnbaum-Saunders-$t$]{\includegraphics[height=4.0cm,width=4.0cm]{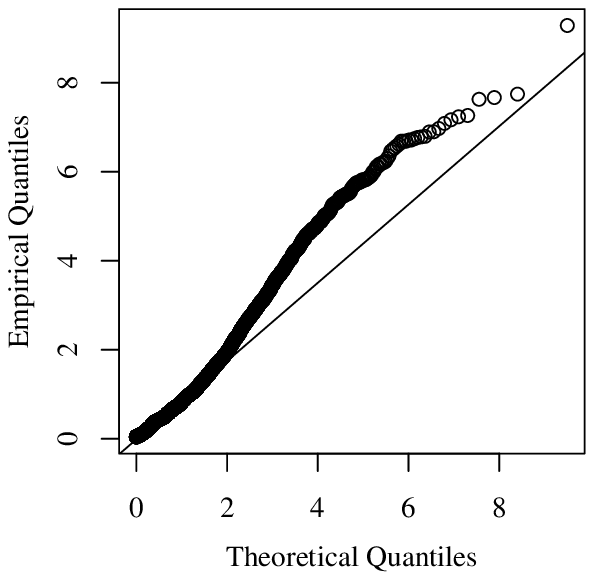}}

 \caption{\small {QQ plot and its envelope for the generalized Cox-Snell residuals in the indicated model for the Apple data ($q=0.50$).}}
\label{fig:qqplots_q050}
\end{figure}

\begin{figure}[!ht]
\centering
\subfigure[Log-normal]{\includegraphics[height=4.0cm,width=4.0cm]{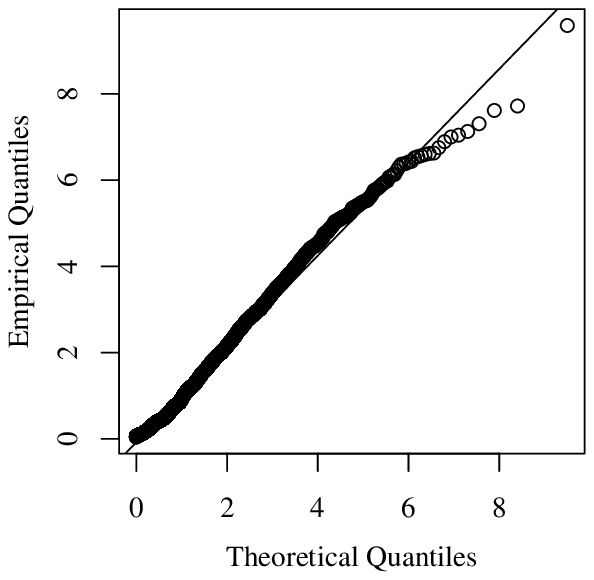}}
\subfigure[Log-Student-$t$]{\includegraphics[height=4.0cm,width=4.0cm]{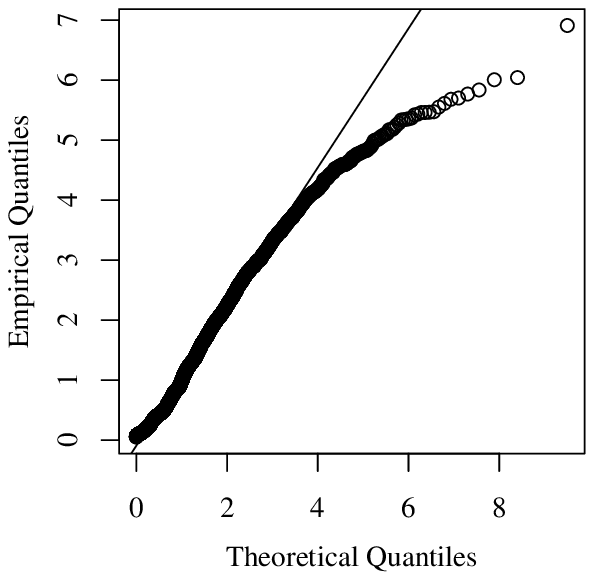}}
\subfigure[Log-power-exponential]{\includegraphics[height=4.0cm,width=4.0cm]{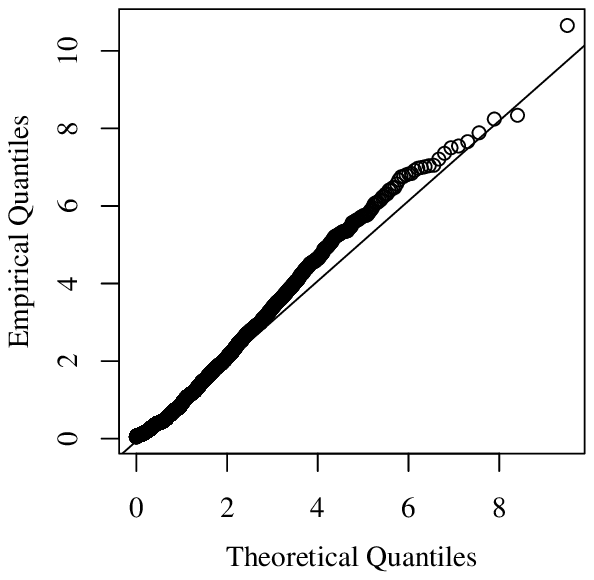}}
\subfigure[Log-hyperbolic]{\includegraphics[height=4.0cm,width=4.0cm]{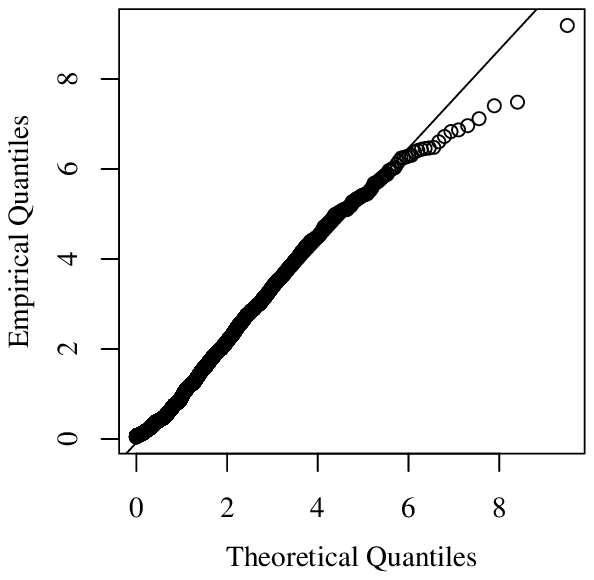}}\\
\subfigure[Log-slash]{\includegraphics[height=4.0cm,width=4.0cm]{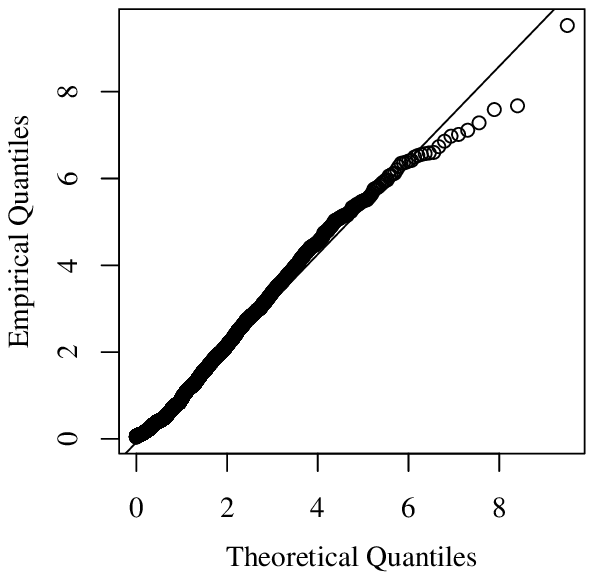}}
\subfigure[Log-contaminated-normal]{\includegraphics[height=4.0cm,width=4.0cm]{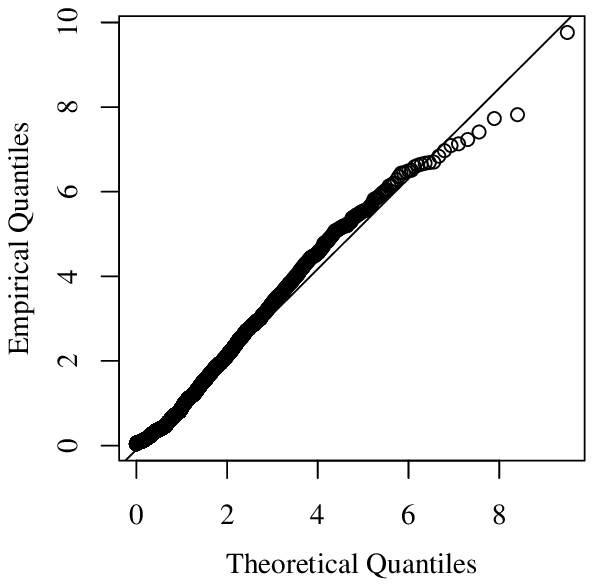}}
\subfigure[Extended Birnbaum-Saunders]{\includegraphics[height=4.0cm,width=4.0cm]{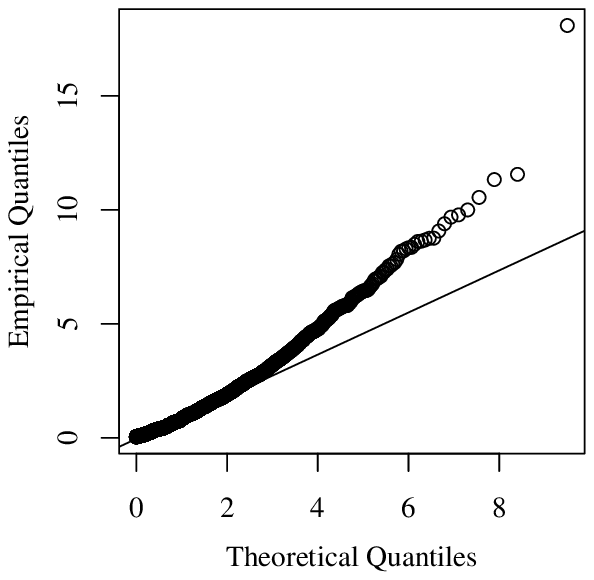}}
\subfigure[Extended Birnbaum-Saunders-$t$]{\includegraphics[height=4.0cm,width=4.0cm]{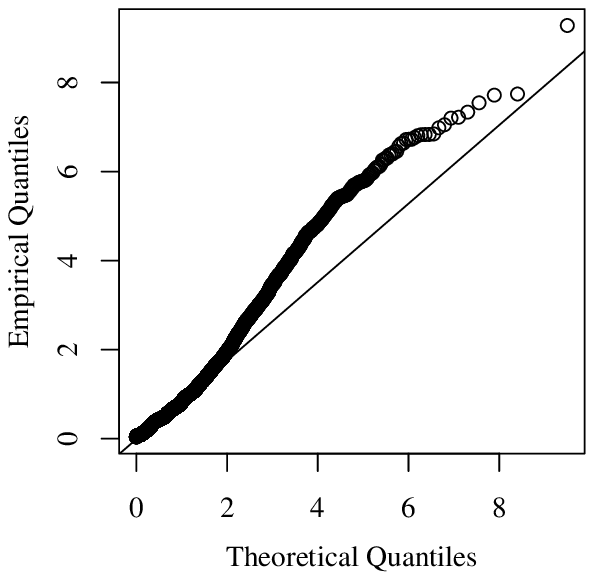}}

 \caption{\small {QQ plot and its envelope for the generalized Cox-Snell residuals in the indicated model for the Apple data ($q=0.975$).}}
\label{fig:qqplots_q0975}
\end{figure}

We first investigate the best model among the class of QLS-ACD models over $q \in \{0.01,0.02,$ $\dots,0.99\}$. To this end, we initially fit the models based on a grid of values of $q \in \{0.01,0.02, \dots,0.99\}$ and then we compute the averages of the
corresponding Akaike (AIC), Bayesian (BIC), corrected Akaike (CAIC), and Hannan-Quinn (HQIC) information criteria values.
From Table \ref{tab:criterions}, we observe that the quantile extended Birnbaum-Saunders-$t$ ACD model provides the best adjustment compared to the other QLS-ACD models based on the values of AIC, BIC, CAIC, and HQIC. However, these values are not substantially different among all models. Figures~\ref{fig:qqplots_q0025}--\ref{fig:qqplots_q0975} show the QQ plots of the GCS residuals for the models considered in Table \ref{tab:criterions} with $q \in \{0.025,0.50,0.975\}$. From these figures, we note that the quantile log-Student-$t$, extended Birnbaum-Saunders and extended Birnbaum-Saunders-$t$ ACD models do not provide good fits.

Based on the results of the information criteria (Table \ref{tab:criterions}) and the QQ plots of the residuals (Figures~\ref{fig:qqplots_q0025}--\ref{fig:qqplots_q0975}), we choose two quantile ACD models: log-normal and log-power-exponential. In the log-normal case, it presents values of information criteria close to those of log-hyperbolic, log-slash and log-contaminated-normal with the advantage of offering parsimony, that is, there is no extra parameter. For the log-power-exponential case, we have the lowest information criteria values after the extended Birnbaum-Saunders-$t$ model, and the QQ plots provide good fits.

\begin{table}[!ht]
	\centering
	\caption{Averages of the AIC, BIC, CAIC, and HQIC values computed across $q \in \{ 0.01,0.02, \ldots, 0.98, 0.99\}$ for the indicated QLS-ACD models with the Apple data.}  \label{tab:criterions}

\begin{tabular}{lccccccccc}
  	\toprule
   Distribution & AIC & BIC & CAIC & HQIC \\
  \midrule
   Log-normal & 11205.24& 11232.48 &11205.25 &11205.95 \\
   Log-Student-$t$ & 11282.27 &11309.50 &11282.28 &11282.97\\
   Log-power-exponential & 11135.92 &11163.15 &11135.93 &11136.62 \\
   Log-hyperbolic & 11218.12 &11245.35 &11218.13 &11218.82 \\
   Log-slash & 11206.49 &11233.73 &11206.50 &11207.20 \\
   Log-contaminated-normal & 11205.81 &11233.05 &11205.82 &11206.52 \\
   Extended Birnbaum-Saunders & 11125.75 &11152.98 &11125.76 &11126.45 \\
   Extended Birnbaum-Saunders-$t$ & 10928.56 & 10955.79  &10928.57 &10929.26 \\
 	\bottomrule
\end{tabular}
\end{table}

\begin{figure}[!ht]
\vspace{-0.25cm}
\centering
{\includegraphics[scale=1]{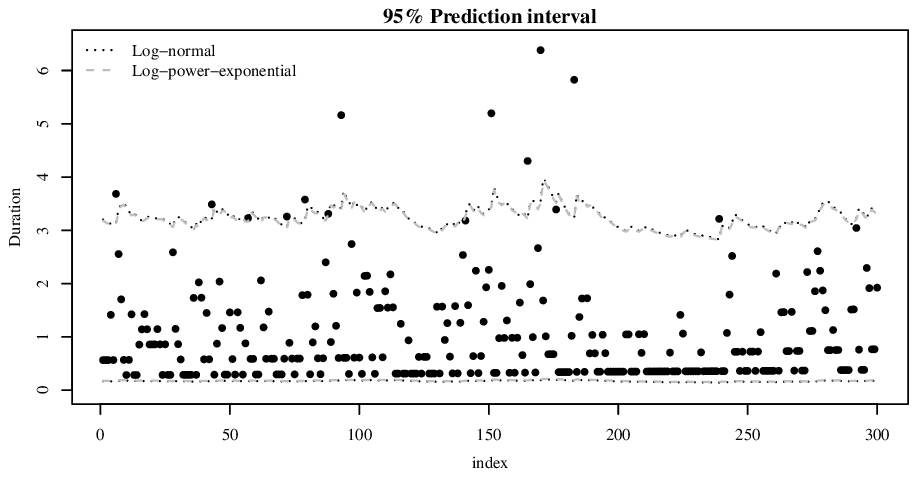}}
\vspace{-0.2cm}
\caption{{95\% prediction intervals (one-step-ahead) from the quantile log-normal and log-power-exponential ACD models with the Apple data.}}\label{figfore:1}
\end{figure}

Figure \ref{figfore:1} shows the plots of 95\% prediction intervals from the quantile log-normal and log-power-exponential ACD models with the Apple data. To construct this plot, we estimate the parameters of the ACD models for each value of $q\in\{0.025,0.975\}$ and then obtain the respective one-step-ahead predictions, which means that the last observation is not included in the estimation. We compute the predictions for the last 300 observations. Figure \ref{figfore:1} shows that 96.33\% and 96\% of the observations are within the limits of the prediction interval for the quantile log-normal and log-power-exponential ACD models, respectively. Therefore, both models provide values close to the nominal level of 95\%.

\subsection{An algorithm for IVaR forecasting}\label{sec:05.3}

We now propose an adaptation of the IVaR measure studied by \cite{cht:07}.  The proposed intraday VaR is a couple $\left(\widehat{\Psi}_{q, n+1},\text{IVaR}_{n+1}(\alpha)\right)$, where $\widehat{\Psi}_{q, n+1}$ provides the forecast of the quantile duration before the occurrence of the next price change, and $\text{IVaR}_{n+1}(\alpha)$ is the forecast of the level of risk such that
$$P\left( r_{\tau_{n+1}} < - \text{IVaR}_{n+1}(\alpha)| \mathcal{F}_{n}\right)=\alpha.$$

The steps involved in the IVaR forecasting are presented below in the form of an algorithm; see Algorithm \ref{alg:ivar}. As \cite{cht:07} explain, the IVaR corresponds to the maximum expected loss that will not be surpassed (at a given confidence level $\alpha$), at the next price event, provided this event takes place. An advantage of the proposed IVaR in Algorithm \ref{alg:ivar} over the original IVaR introduced by \cite{cht:07}, is that it allows us to assess different percentiles instead of the traditionally used conditional mean duration. Therefore, the level of risk can be assessed by taking into account directly predicted extreme quantile durations, for example.

\begin{algorithm}[!ht]
\floatname{algorithm}{Algorithm}
\caption{Steps involved in the IVaR forecasting.}\label{alg:ivar}
\begin{algorithmic}[1]
\State Given the seasonally adjusted price durations, $\{\tilde{x}_t\}_{t=1}^{n}$, choose an appropriate QLS-ACD($r,s,q$) model and retrieve the corresponding estimates of the model parameters $\boldsymbol{\theta}$. Compute the one-step-ahead out-of-sample quantile duration for the price change number $n+1$ as follows:
\begin{eqnarray}\label{conditional_quantile_forecast}
h(\widehat{\Psi}_{q, n+1}) = \widehat{\omega} + \sum_{j=1}^{r}\widehat{\alpha}_j \text{log}(\widehat{\Psi}_{q, n+1-j}) + \sum_{j=1}^{s}\widehat{\beta}_j \left(\frac{x_{n+1-j}}{\widehat{\Psi}_{q, n+1-j}}\right).
\end{eqnarray}

\State Based on the QLS-ACD($r,s,q$) model parameter estimates obtained in Step 1, compute the forecast value of conditional instantaneous intraday volatility as
\begin{eqnarray}
 \widehat{\sigma}^2(\tau_{n+1}|\mathcal{F}_{n})=\widehat{h}_{X_{n}|\mathcal{F}_{n}}(x_{n};\widehat{\Psi}_{q,n}, \widehat{\phi})\left(\frac{\kappa}{p_{\tau_n}} \right)^2\frac{1}{\widehat{\varpi}_{n}},
\end{eqnarray}
where $\widehat{h}_{X_{n}|\mathcal{F}_{n}}(x_{n};\widehat{\Psi}_{q,n}, \widehat{\phi})$ is the prediction of conditional hazard function for the price durations, $\kappa$ is the magnitude of the price change, $p_{\tau_n}$ is the mid-point of the bid and ask prices, and $\widehat{\varpi}_n$ is the intraday seasonality.

\State Based on the parameter estimates from Step 2, compute the in-sample volatilities as
\begin{eqnarray}
 \widehat{\sigma}^2(\tau_{t}|\mathcal{F}_{t-1})=\widehat{h}_{X_{t}|\mathcal{F}_{t-1}}(x_{t};\widehat{\Psi}_{q,t}, \widehat{\phi})\left(\frac{\kappa}{p_{\tau_{t-1}}} \right)^2\frac{1}{\widehat{\varpi}_{t-1}},\quad t=1,\ldots,n.
\end{eqnarray}
Compute also the empirical $\alpha$-quantile of the standardized series of returns
$$\widehat{\epsilon}_{\tau_t}=\frac{{r}_{\tau_t}}{\sqrt{\widehat{\sigma}^2(\tau_{t}|\mathcal{F}_{t-1})}},\quad t=1,\ldots,n,$$
denoted by $q_{\alpha}(\widehat{\epsilon})$.

\State Compute the $\text{IVaR}_{n+1}(\alpha)$ as follows:
$$\text{IVaR}_{n+1}(\alpha)|\mathcal{F}_{n}=q_{\alpha}(\widehat{\epsilon})\,\sqrt{\widehat{\sigma}^2(\tau_{n+1}|\mathcal{F}_{n})}.$$

\State  The proposed intraday VaR is then the couple $\left(\widehat{\Psi}_{q, n+1},\text{IVaR}_{n+1}(\alpha)\right)$, where $\widehat{\Psi}_{q, n+1}$ and $\text{IVaR}_{n+1}(\alpha)$ are obtained from Step 1 and 4, respectively.
\end{algorithmic}
\end{algorithm}

\begin{figure}[!ht]
\centering
\subfigure[Log-normal]{\includegraphics[scale=1]{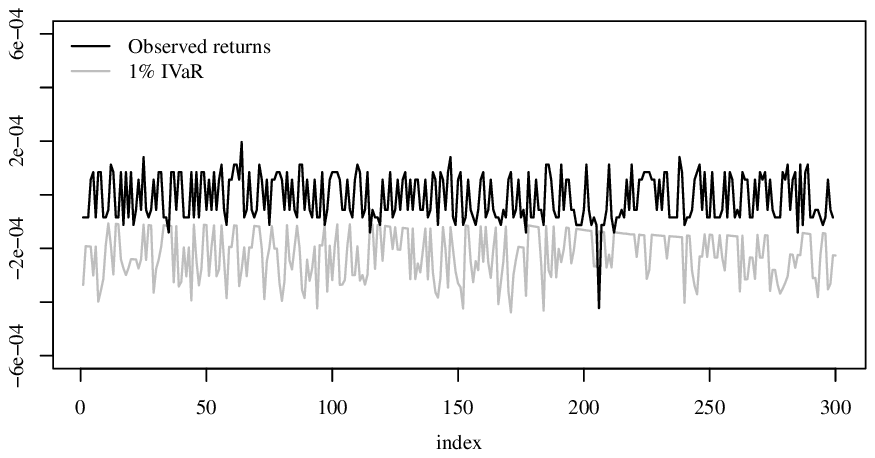}}
\subfigure[Log-power-exponential]{\includegraphics[scale=1]{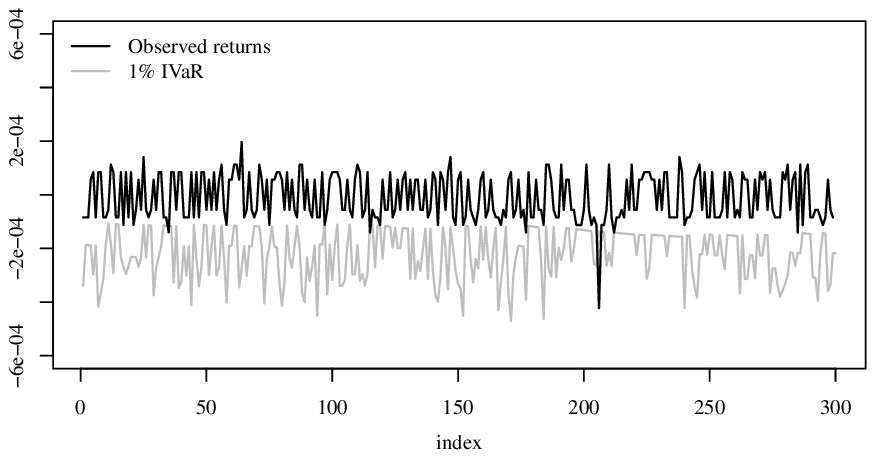}}
 \caption{\small {Out-of-sample forecasts of 1\% IVaR and observed returns for the Apple data.}}
\label{fig:qqplots_ivar}
\end{figure}

\begin{figure}[!ht]
\centering
{\includegraphics[scale=1]{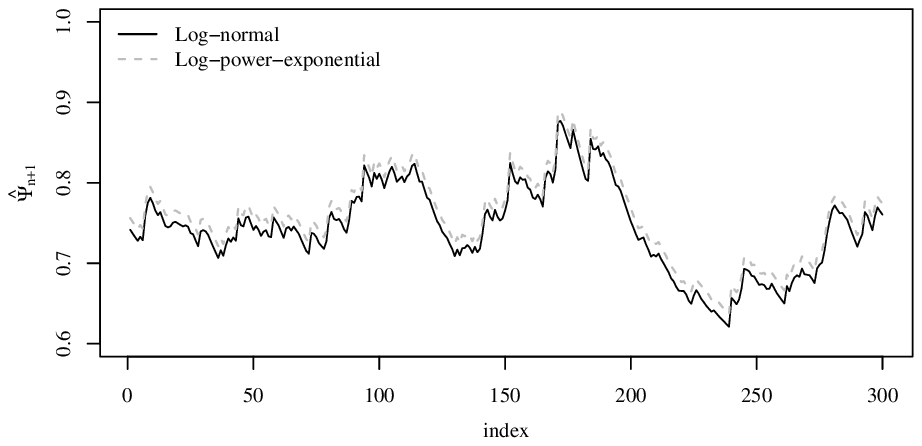}}
 \caption{\small {Out-of-sample forecasts of median quantile ($\widehat{\Psi}_{q=0.50, n+1}$) for the Apple data.}}
\label{fig:qqplots_median}
\end{figure}

We run Algorithm \ref{alg:ivar} with the quantile log-normal and log-power-exponential ACD models ($q=0.50$ and $\alpha=1\%$). Figures \ref{fig:qqplots_ivar} and \ref{fig:qqplots_median} show the 1\% out-of-sample IVaR with the corresponding returns of price events and the out-of-sample median durations for the price change from the quantile log-normal and log-power-exponential ACD models with the Apple data. From Figure \ref{fig:qqplots_ivar}, we observe that the evolution of price events returns is tracked quite well by the IVaR forecasts. In terms of hits, which is defined by
$$
\bar{\mathcal{I}}=\frac{1}{\mathcal{T}}\sum_{t=1}^{\mathcal{T}}\mathcal{I}_t,
$$
with
$$
\mathcal{I}_t=
\begin{cases}
    1   & \quad \text{if } {r}_{\tau_t} <\text{IVaR}_{t}(\alpha),\\
    0  & \quad \text{if } {r}_{\tau_t} \geq \text{IVaR}_{t}(\alpha),
  \end{cases}
$$
the quantile log-power-exponential ACD model takes advantage with $\bar{\mathcal{I}}=1.33\%$ over the quantile log-normal ACD model with $\bar{\mathcal{I}}=1.67\%$.

\clearpage

\section{Concluding remarks}\label{sec:06}

In this paper, we have proposed quantile autoregressive conditional duration models based on log-symmetric distributions. The proposed models are defined in terms of a conditional quantile duration that facilitates, for example, the computation of prediction intervals. A Monte Carlo simulation has been carried out for evaluating the performance of the maximum likelihood estimates and for the evaluation of the empirical distribution of the generalized quantile residuals. We have applied the proposed models to a real financial duration data set corresponding to price durations of Apple Inc. stock. The results support the suitability of the proposed class of quantile-based log-symmetric ACD models, both in terms of model fitting and forecasting ability. We have also derived a semi-parametric intraday value-at-risk model based on the proposed ACD models. The intraday value-at-risk model tracked well the evolution of price change returns. The results of this paper can be very useful for active market participants such as high frequency traders and financial institutions. As part of future research, it will be of interest to propose bivariate models along the lines of \cite{mosconiolivetti:05}. Furthermore, a test for the predictive abilities of the proposed intraday value-at-risk model can be developed \citep{cht:07}. Work on these problems is currently in progress and we hope to report these findings in future manuscripts.\\



\end{document}